\documentclass[aps,prl,epsfig,twocolumn,showpacs]{revtex4-1}

\usepackage{epsfig,amssymb,amsmath,latexsym} 
\usepackage{subfigure}
\usepackage{wrapfig}
\usepackage{float}
\usepackage{color}
\usepackage{bm,bbm,dsfont}
\usepackage[colorlinks=true,linktoc=page,linkcolor=magenta,citecolor=magenta]{hyperref}

\newcommand\so{{{SOI}}}
\newcommand\ie{{\it{i.e.}}}

\newcommand\beq{\begin{equation}}
\newcommand\eeq{\end{equation}}

\newcommand\bea{\begin{eqnarray}}
\newcommand\eea{\end{eqnarray}}

\newcommand\bi{\begin{itemize}}
\newcommand\ei{\end{itemize}}

\newcommand\kso{k_{\rm so}}

\def\tr{\mathop{\rm tr}\nolimits}

\newif\ifboo \boofalse

\begin{document}

\textheight=23.8cm

\title{Transport signature of fractional Fermions in Rashba nanowires}
\author{Diego Rainis, Arijit Saha, Jelena Klinovaja, Luka Trifunovic, and Daniel Loss}
\affiliation{
\mbox{Department of Physics, University of Basel, Klingelbergstrasse 82, CH-4056 Basel, Switzerland}\\
}

\date{\today}

\begin{abstract}
We study theoretically transport through a  semiconducting nanowire (NW) in the presence of Rashba spin orbit interaction, uniform magnetic field, and spatially  modulated magnetic field. 
The system is fully gapped, and the interplay between the spin orbit interaction and the magnetic fields leads to fractionally charged fermion (FF) bound states of Jackiw-Rebbi type at each end of the nanowire. 
We investigate the transport and noise behavior of a N/NW/N system, where the wire is contacted by two normal leads (N), and we look for possible signatures that could help in the experimental detection of such states. 
We find that the differential conductance and the shot noise exhibit a sub-gap structure which fully reveals the presence of the FF state. 
Our predictions can be tested in standard two-terminal measurements through InSb/InAs nanowires.
\end{abstract}

\maketitle

{\it Introduction.~}%
In the last years considerable attention has been devoted to  condensed-matter systems where exotic fractionally charged excitations 
form, which are interesting both from a fundamental point of view and for quantum computation purposes~\cite{ady1}.
Very recently, a proximity effect involving fractionally charged edge states has
been considered~\cite{Stern_Parafermions_PRX,Alicea_Parafermions_NatComm,
Cheng_Parafermions,Vaezi_Parafermions,Ady_Stern,11_authors}, where the induced zero modes inherit a fractional exchange 
phase giving rise to parafermions. 

Another setup that admits fractional-charge excitations,
with peculiar bound states of Jackiw-Rebbi type ~\cite{jackiewrebbi},
involves finite-length NWs and rings in the presence of a charge-density-wave gap induced by a periodic modulation of the chemical potential~\cite{suhas1} and in a 
quantum spin Hall state induced by a magnetic domain wall~\cite{sczhang}.
 Similarly, it was shown that the simultaneous presence of Rashba spin-orbit (SOI) interaction and uniform and spatially periodic magnetic field can produce gapped phases 
with bound states and a significantly richer phase diagram~\cite{jelena1}. This includes a reentrance behavior of Majorana fermions (MFs) and a new phase characterized by 
fractionally charged fermions (FF), analogous to Jackiw-Rebbi fermions of charge $e/2$~\cite{jackiewrebbi,rajaramanbell} or fractional charge formation in the SSH 
model with long-chain polyenes~\cite{ssh1,ssh2}. Unlike MFs, the FF states exist both with and without superconductivity, and can exhibit non-abelian braiding statistics~\cite{jelena3}.

\begin{figure}[h!]
\begin{center}
\includegraphics[width=0.90\linewidth]{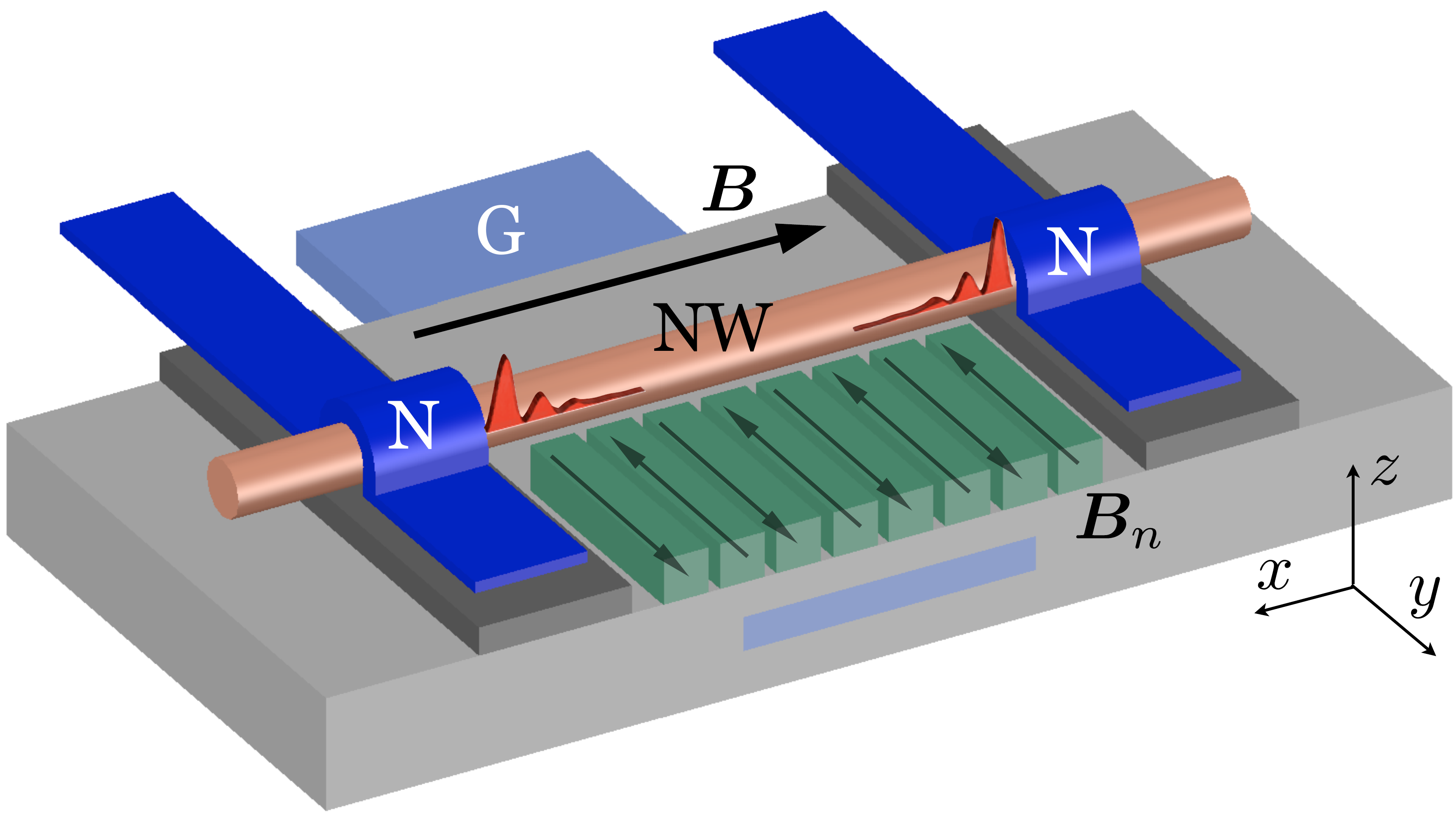}
\caption{Sketch of the transport setup, consisting of a  semiconducting nanowire NW (pink) of length $L$ attached to normal N leads (blue). The chemical potential in the NW is controlled by the gate G (light blue).
A uniform magnetic field ${\bm {B}}$ is applied along the wire.
The NW is also subjected to a spatially varying magnetic field ${\bm {B}_{n}(x)}$ produced by periodically arranged nanomagnets (green).
A fractional-fermion bound state forms at each wire end (red). 
}
\label{figgeometry}
\end{center}
\end{figure}

However, this rich phenomenology notwithstanding, the transport and noise signature of FFs in these models have not been investigated  so far.
Therefore, in this Letter we address the question of finding  transport signatures of the FFs in  non-superconducting Rashba NWs, which we regard as one of the most promising setups for the observation of fractional fermions.
To carry out our analysis, we consider the NW contacted by two normal (N)
leads at its two ends (see Fig.~\ref{figgeometry}). 
We adopt the standard scattering formalism~\cite{Lambert1991JPCM_Landauer,Lambert1993JPCM,Lambert1998JPCM_Raimondi,ButtikerBlanter} to find the differential conductance and the 
shot noise in the presence of the FFs state.
We find that the FF phase is identified in a distinctive way by a series of features in the conductance behavior of the junction, especially if spin-sensitive measurements are possible. 
The fractional charge of the bound states cannot however be probed by these transport studies.

{\it Model.~}%
We consider a semiconducting NW of length $L$ along the $\hat x$ direction, in the presence of \so along the $\hat z$ direction and a magnetic field
that includes a uniform ($\bm B$) and a spatially periodic component ($\bm B_n$). 
The NW continuum Hamiltonian is given by
$H_0= \frac{1}{2} \int {\rm d}x \Psi^\dagger (x) \mathcal{H}_0 \Psi (x)$, where 
$\Psi =  (\Psi_\uparrow, \Psi_\downarrow)^{\rm T}$,  $\Psi_{\sigma}(x)$ is the annihilation  operator for a spin-$\sigma$ electron at position $x$. The Hamiltonian density is of the form
\bea
\mathcal{H}_0& =- \hbar^2 \partial_x ^2/2m-\mu - i \alpha  \sigma_3  \partial_x   \ ,
\label{h_0_density}
\eea
with $m$ the electron band mass, $\alpha$ the \so coefficient, and $\mu$ the chemical potential. 
The spectrum of $\mathcal{H}_0$ consists of two parabolas centered at the SOI momenta $\pm k_{\rm so}=\pm m\alpha/\hbar^2$.
%
The magnetic field leads to the Zeeman term
\bea
\mathcal{H}_{z}= g\mu_{\rm B}  [{\bm B}+{\bm B}_{n}(x)]  \cdot {\boldsymbol \sigma}\ /2 \ ,
\label{hzeeman}
\eea
where $g$ is the Land\'e g-factor and $\mu_B$ the Bohr magneton. 
$\bm B$, chosen along the $\hat x$ direction, opens a gap of magnitude $\Delta_z=g \mu_{\rm B} B/2$ at $k=0$.
The oscillating field ${\bm B}_{n}(x)$ is oriented along $\hat y$, 
${\bm B}_{n}={\hat y}B_{n}\sin (4 k_{\rm so} x+\theta)$ (but other equivalent configurations are possible~\cite{jelena1}).
It couples the two large-$k$ branches of the spectrum,
and opens up a gap of magnitude $\Delta_n=g \mu_{\rm B} B_n/4$. It can be generated externally, by an array of nanomagnets placed in proximity 
to the wire~\cite{magnets_Flensberg,PRX_Daniel,Karmakar_Nanomagnets}, or internally, {\it e.g.}, by the hyperfine field of ordered nuclear spins inside the nanowire~\cite{daniel1}.

\begin{figure}[h]
  \centering
{\includegraphics[width=0.80\linewidth]{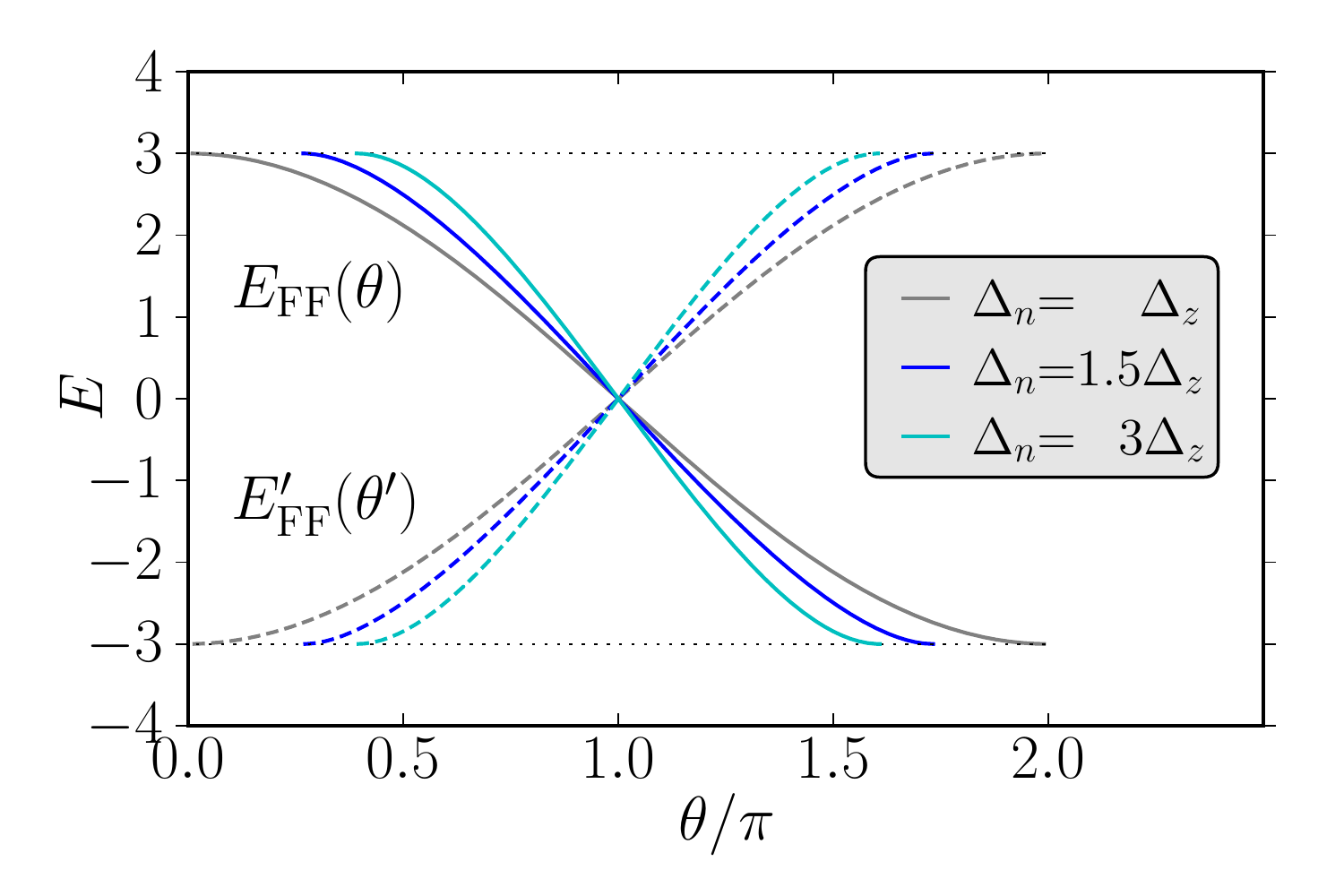}}
   \caption{Spectra $E_{\rm FF}(\theta)$ and $E_{\rm FF}^\prime(\theta^\prime)$ of the left (full lines) and right (dashed lines) FF bound state, for $L\rightarrow\infty$, for different gap values, as evaluated from the analytical result Eq.~(\ref{ff_disp}). Here we chose symmetric left and right boundary conditions ($\theta^\prime=\theta+n\pi$).
   The bound state energy vanishes for $\theta=\pi$, and becomes for $\theta$ away from it, merging into the continuum at $E=\min\{\Delta_z,\Delta_n\}$, in correspondence of $\bar\theta=\bar\theta(\Delta_z/\Delta_n)$.}
\label{fig:bulk_spectrum}
\end{figure}

In our analysis, we are assuming that the \so energy $m \alpha^2/ \hbar^2$ is the largest energy scale in the NW. 
In this strong-\so regime, we follow the procedure described in Refs.~[\onlinecite{jelena4,jelena2}], which allows us to linearize $\mathcal{H}_0$ around $k=0$ (referred to as interior branch) and $k=\pm 2 k_{\rm so}$ (referred to as exterior branches).
%
Finally, for $\mu=0$, one obtains the spectrum  around $k=0$ and $k=\pm2k_{\rm so}$ as 
$E_{m}^2=(\alpha k)^2+\Delta_m^2 \ ,$
with a gap for the  interior ($m=i$) and exterior ($m=e$) branch given by $\Delta_z$ and $\Delta_n$, respectively.
%
In the presence of such two magnetic fields, the system is fully gapped, with no propagating modes at subgap energies, $|E|<\min\{\Delta_z,\Delta_n\}$. However, as shown in Ref.~[\onlinecite{jelena1}], there can be localized edge states, FFs, in the gap. For example, in a semi-infinite geometry there is one bound state, localized at the end of the nanowire, with energy given by
\bea\label{ff_disp}
E_{\rm FF} = \frac{\Delta_z \Delta_n \sin \theta}{\sqrt{\Delta_z^2 + \Delta_n^2 -2\Delta_z\Delta_n \cos \theta}}\;.
\eea
The angle $\theta$ encodes the boundary condition for the oscillating field at the nanowire edge.
The FF state exists only if the following relation  is satisfied: $ \cos \theta<\min\{\Delta_z,\Delta_n\}/\max\{ \Delta_z,\Delta_n\} \leq 1$, see Fig.~\ref{fig:bulk_spectrum}.

\begin{figure}
\begin{center}
\includegraphics[width=1.0\linewidth]{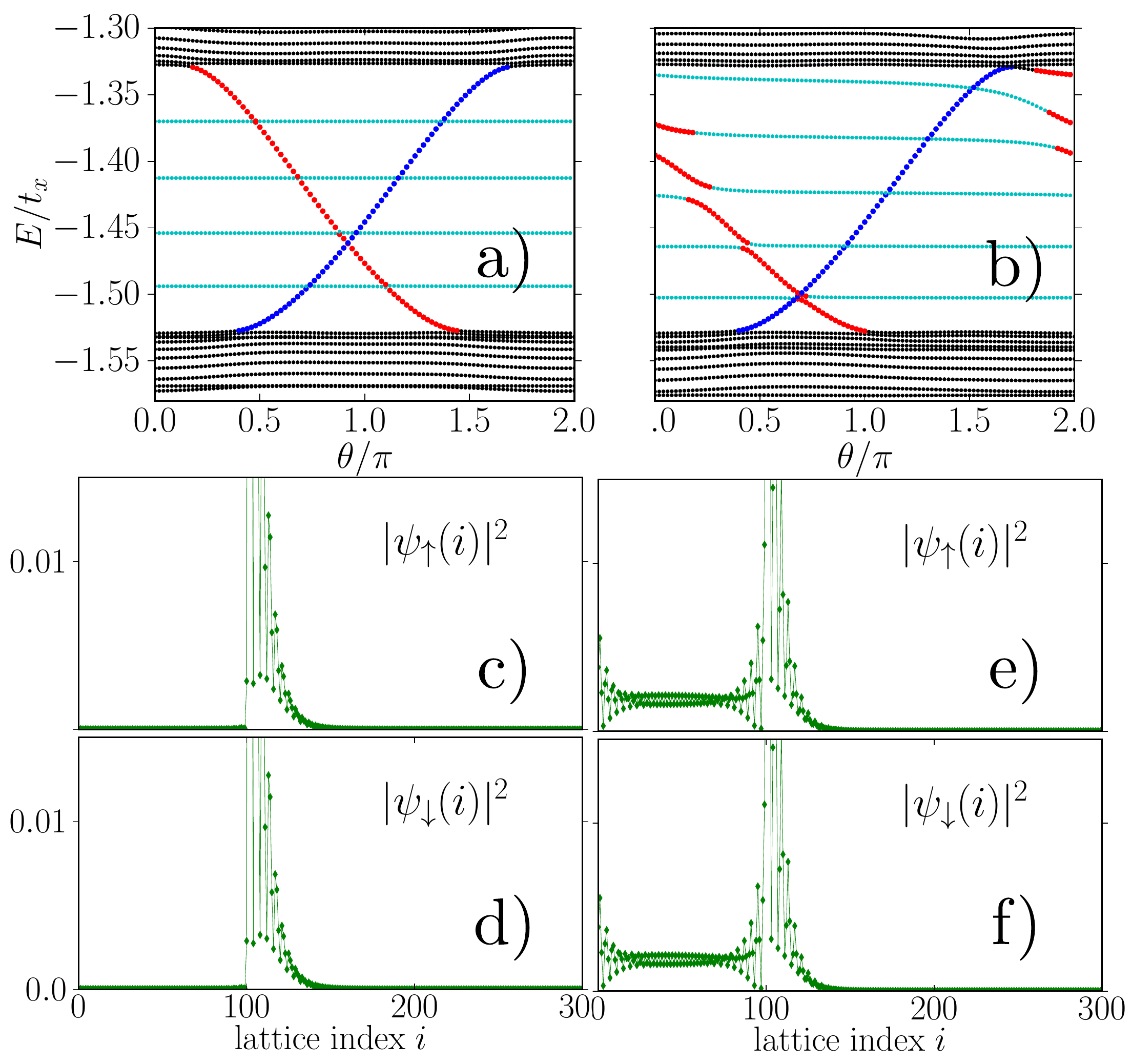}
\caption{Tight-binding numerical results for a nanowire with a normal/fractional-fermion (N/FF) junction, with FF bound states in the FF phase. 
$\Delta_z/t_x=0.2$ and $\Delta_n/t_x=0.1$ ($t_x$ hopping parameter).
The rotating field $B_n$ is present only in the FF section of length $L_{\rm T}$, with $L_{\rm T}=3L_{\rm N}$ and $L_{\rm N}=100$. 
At the N-FF interface there is a rectangular barrier of width $L_b=L_{\rm T}/50$, and height $V/t_x =5$ (left column) or $V/t_x =1$ (right column). 
a)-b): TB energy spectrum $E(\theta)$. Inside the bulk gap, delimited by continuum states (black dots), there are two FF bound states, localized at the left (red dots) and 
right end (blue dots). In addition, there are states localized in the N section (cyan dots).
c)-f): Probability densities $|\psi_{\sigma}(x)|^2$ of the left FF state at $\theta=0.6\pi$ [c),d)] and $\theta=0.1\pi$ [e),f)].
If the barrier is high, the FF is completely localized in the FF section [c),d)], and, as a result, the FF spectrum shows no hybridization [a)] and is fully consistent 
with Eq.~(\ref{ff_disp}). For lower barrier,
the left FF leaks out in the normal section [e),f)] and gets hybridized with the   N-region subgap states [b)].}
\label{tight_binding_spectrum}
\end{center}
\end{figure}

The non-interacting, spin-degenerate left ($l=$L) and right ($l=$R) leads are described by the Hamiltonian density
$
\mathcal{H}_{l}=- \hbar^2 \partial_x ^2/2m  - \mu_{l} \;.
\label{hleads}
$
At each lead/wire interface we insert a barrier, modeled with a $\delta$-function in the wave-function-matching analysis and with a rectangular-shaped potential in the tight-binding (TB) calculations.
In the latter, more realistic case, one observes that if the barrier between NW and N sections is high enough the FFs are localized entirely in the NW section, see Figs.~\ref{tight_binding_spectrum}c)-d), and the spectra of NW and N section are decoupled, as in Fig.~\ref{tight_binding_spectrum}a). 
However, if the barrier is reduced, the left FF wave function (WF) leaks out in the N region and hybridizes with the local WFs, see Figs.~\ref{tight_binding_spectrum}e)-f) and the red spectrum in Fig.~\ref{tight_binding_spectrum}b).

{\it Method.~}%
To study the transport through the semiconducting NW of length $L$ we employ an $\mathbbm S$-matrix formalism,
where we match wave-functions and associated currents at each wire/lead interface~\cite{rashba1,rashba2,rashba3}. 
As main transport signal we then consider the transmission probability,  which at zero temperature gives the differential conductance (${dI}/{dV}$) in units of $e^2/h$.
Depending on the choice of the spin basis in the leads, one observes different types of behavior in the spin-resolved transport coefficients ${\mathbbm S}_{\sigma\sigma^\prime}$.	
With the given configuration of SOI and magnetic fields, 
the most natural choice (provided by the eigen-basis of $\mathbbm {t^\dagger t}$) is to work in the $y$ basis.
%
Therefore, we consider incoming plane-wave WFs 
with wave vector $k_{l}=\sqrt{2m(\mu_l+E)/\hbar^{2}}$, and
%
a nanowire WF that in the strong-\so limit reads
$\psi_{\rm NW}= a\psi_{  i}^{  +} + b\psi_{  i}^{-} + c\psi_{  e}^{+}+ d\psi_{  e}^{-}  \;,$
%
%
where $\psi_{  i/e}^{\pm}$ are the four solutions (internal/external branches, right/left movers) at energy $E$, given by
\begin{align}\label{erashba}
\psi_{  i}^{\pm}(x)&= \left( {\textstyle \frac{\left[E\pm\sqrt{E^{2}-\Delta_{z}^{2}}\right]}{\Delta_{z}} } ,~ 1 
\right){e}^{\pm i k_z x} \;,  \\
\psi_{  e}^{\pm}(x)&=\left( {\textstyle \frac{{e}^{i\theta}\left[E\mp\sqrt{E^{2}-\Delta_{n}^{2}}\right]}{\Delta_{n}} }  
{e}^{-2i\kso x} ,~ {e}^{2i \kso x} \right){e}^{\pm i k_n x}  \;,  \nonumber 
\end{align}
where  $k_z={\sqrt{E^{2}-\Delta_{z}^{2}}}/{\alpha}$ and $k_n={\sqrt{E^{2}-\Delta_{n}^{2}}}/{\alpha}$ are the wave vectors at energy $E$ associated to internal and external branches ~\cite{jelena1}.
The definitions for $k_z$ and $k_n$ apply both above and below the gap, where they become purely imaginary and describe evanescent modes, whose linear combination gives rise to our FF bound states.

By matching wave functions and probability currents at the two N/NW  interfaces
($x=0,L$), we obtain eight linear equations, whose solution are the transport coefficients entering the   $\mathbbm S$-matrix.
%
These quantities all depend on the incoming energy $E$, on the wire length $L$, on $\Delta_{z}$, $\Delta_{n}$, on the phase $\theta$, and on the strengths 
$\lambda_{1}$, $\lambda_{2}$ of the two $\delta$-function barriers at $x=0$ and $x=L$, respectively.

{\it Results.~}%
Here we present the results of our numerical analysis on transport through the  SOI-coupled NW. 
For finite wire length, the two FF end states have a finite overlap, determined by the  localization length of the FFs, which is set by the two gaps, $\xi_{z}(E)\simeq \alpha/\sqrt{\Delta_{z}^2-E^2}$ and $\xi_{n}(E)\simeq \alpha/\sqrt{\Delta_{n}^2-E^2}$.
%
For long wires, $L\gg\xi_z,\xi_n$, the two FF states are  decoupled, and each of them has an energy approximately given by the semi-infinite geometry result Eq.~(\ref{ff_disp}). Such ``ideal'' spectrum is plotted in Fig.~\ref{fig:bulk_spectrum}, where we show the energies $E_{\rm FF}(\theta)$ and $E_{\rm FF}^\prime(\theta^\prime)$ of the left and right bound states ($\theta^\prime=\theta+4k_{\rm so}L$ encodes the boundary condition of $B_n$ at the right end).  

\begin{figure}[h]
  \centering
{\includegraphics[width=1\linewidth]{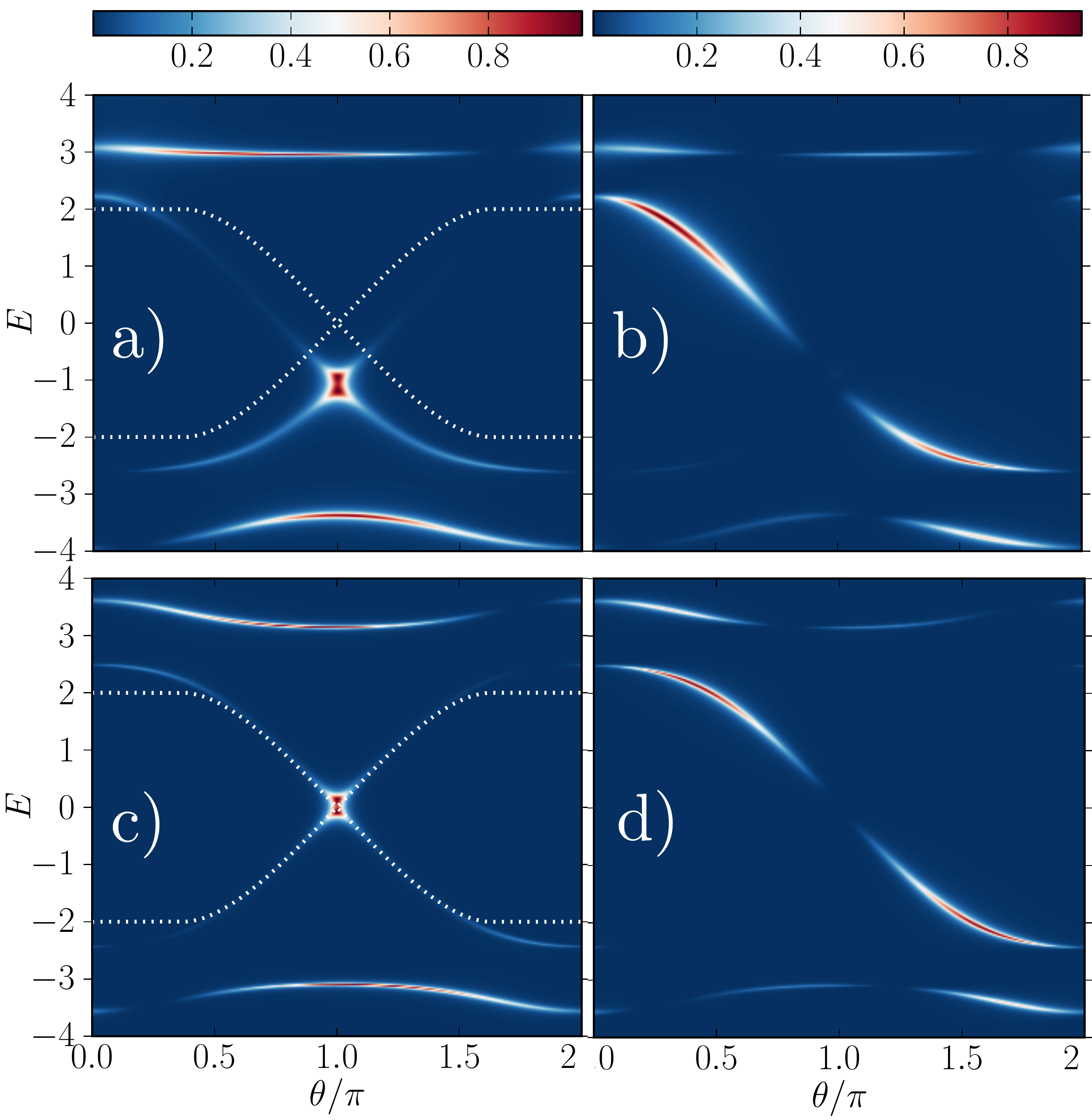}}
   \caption{The coefficients $T_\uparrow=T_{\uparrow\uparrow}+T_{\downarrow\uparrow}$ [a) and c)] and $R_{\uparrow\downarrow}$ [b) and d)] are plotted in the $E$-$\theta$ plane,  as evaluated in two different regimes.
   Panels a) and b) refer to the high-barrier ($\lambda_1$=$\lambda_2$=$10\alpha$), small-$\mu_l$ limit ($k_l\simeq k_{\rm so}/2$).
Panels c) and d) are instead obtained in the zero-barrier, large-$\mu_l$ limit ($k_l\simeq 100k_{\rm so}$), which yields an equivalent degree of decoupling, but does not shift the bound state spectra, as highlighted by the white dotted lines, which correspond to Eq.~(\ref{ff_disp}).
 The other  parameter values are $\Delta_{z}=\alpha^2/4$, $\Delta_{n}=\alpha^2/10$, $L\simeq 3\xi_z(0)$.
}
\label{fig:LongWire}
\end{figure}

In Figs.~\ref{fig:LongWire}a) and~\ref{fig:LongWire}b)  we show that the transmission behavior in the long-wire regime exactly follows the dispersion $E_{\rm FF}(\theta)$, provided a suitable parameter configuration is chosen.
More precisely, in order to probe in transport the unperturbed wire spectrum one needs, as usual, to operate in the tunnel regime, \ie, to have small wire-leads coupling. 
This is usually implemented by adding a strong potential barrier (delta-function in the analytics) at the N/NW interfaces.
However, this method has the drawback of modifying the spectra of the bound states, proportionally to the wire-lead coupling strength, see  Figs.~\ref{fig:LongWire}a) and \ref{fig:LongWire}b).
There is another method which allows one to make the bound state signature visible, without the need of introducing a potential barrier at the interfaces.
It amounts to choosing very different chemical potentials (and hence different wave vectors) in the wire and in the leads. The momentum mismatch has then a filtering effect, and the intrinsic properties of the wire are probed without altering the $E(\theta)$ dispersion (only the signal broadening changes, decreasing for larger momentum mismatches), see Figs.~\ref{fig:LongWire}c) and \ref{fig:LongWire}d).

We stress that it is necessary to tune $\mu_l$ in order to make the bound state signatures visible in  ${dI}/{dV}$. For a generic value of $\mu_l$ the momentum filtering acts either ineffectively or too effectively, erasing also the signature of the FF bound state. 

In Fig.~\ref{fig:LongWire}, which contains the main result of this paper, we also show the behavior of the spin-flip reflection coefficient $R_{\uparrow\downarrow}$ as a function of energy and angle $\theta$, in panels b) and d). 
One immediately notices that 
$R_{\uparrow\downarrow}$ exhibits a striking and evident signature of the left bound state, with a peak that almost perfectly traces the left dispersion $E_{\rm FF}(\theta)$. 
Quite intuitively, no influence of the right bound state appears in the behavior of $R_{\uparrow\downarrow}$, apart from at energies very close to the gap, where the localization length increases significantly, and from the anticrossing at small energies, where the two dispersions hybridize and $R_{\uparrow\downarrow}\simeq0$ (consistent with the fact that  at those points $T_\uparrow$ is enhanced to 1).
$R_{\uparrow\uparrow}$ has the complementary behavior (not shown): it is equal to 1 everywhere apart from the energies matching the bound state energy. 
Therefore, if one is able to spin-polarize the incoming current and is able to separately measure $R_{\uparrow\uparrow}$ and $R_{\uparrow\downarrow}$, one should obtain a clear signature of the FF bound state when its energy is matched by the applied bias. 
The additional possibility of tuning $\theta$ ({\it i.e.}~by moving the FF section of the wire) would grant the access to the full $E_{\rm FF}(\theta)$ behavior, an even stronger confirmation for the presence of the FF bound states.

If the polarization axis in the leads is chosen differently, the situation changes slightly:
working  in the $z$ basis, we found that $T_{\uparrow}$ exhibits the same pattern as in the present treatment, but with a maximum value of 1/2. That is due to the spin-projection effect along the $y$ axis that the wire exerts on the incoming electrons. 


Further experimental identification of the FF bound states is possible via noise measurements.
We calculated the shot noise $S\propto \tr\left(  r^\dagger  r  t^\dagger t\right)$  and the  Fano factor $F\propto S/I$ , both for spin-polarized and spin-unpolarized transport.
\begin{figure}[h]
  \centering
{\includegraphics[width=0.7\linewidth]{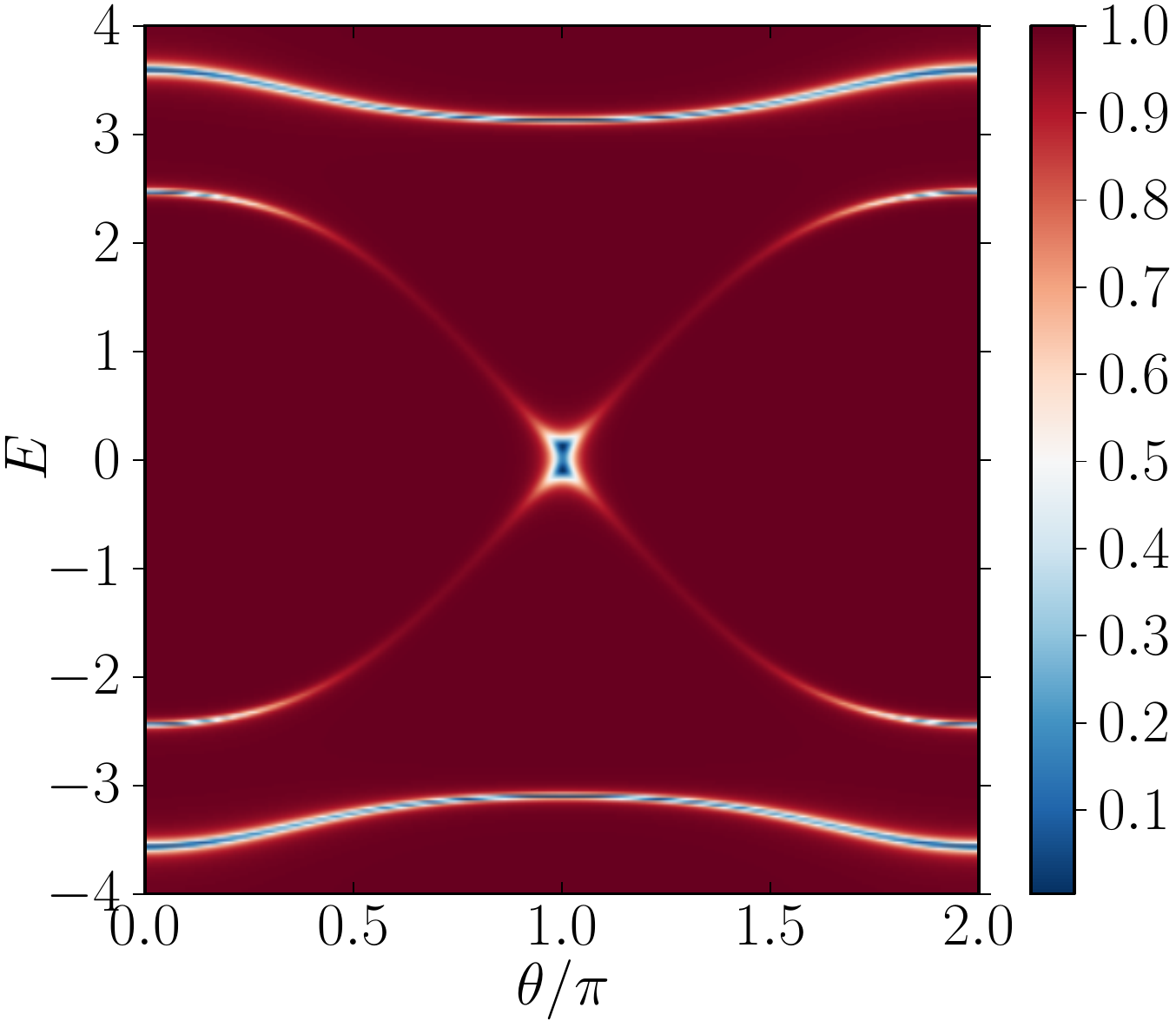}}
   \caption{ 
The Fano factor $F$ for two-terminal spin-unpolarized transport across a nanowire in the FF phase. When the bound state energy is matched, $F$ exhibits a dip with strength given by $T_\uparrow$, signaling transport through the wire. At $E\simeq0$ and $\theta\simeq\pi$ both bound states are accessible and $T_\uparrow\simeq1$, $F=0$. Away from that, transport is impeded and $F=1$. The parameters are the same as in Figs.~\ref{fig:LongWire}c)-d).
}
\label{fig:Fano}
\end{figure}
%
Let us denote $T_\sigma=\sum_{\sigma^\prime}T_{\sigma\sigma^\prime}$.
In the spin-polarized case, one finds $F=1-T_\sigma$, which is the natural generalization of the single-channel formula $F=1-T$.
In the spin-unpolarized case, instead, the interplay of the two spin channels leads to the formula~\cite{ButtikerBlanter}
\begin{align}\nonumber
 F=&\bigg( \sum_{\sigma}T_{\sigma}\bigg)^{-1} \bigg\{ \sum_\sigma \big(  1-T_{\sigma}\big)T_{\sigma}   \\  
&+2\Re e\left[ \left(r^*_{\uparrow\uparrow}r_{\downarrow\uparrow}+r^*_{\uparrow\downarrow}r_{\downarrow\downarrow}\right) 
\left( t^*_{\downarrow\uparrow}t_{\uparrow\uparrow}+t^*_{\downarrow\downarrow}t_{\uparrow\downarrow}  \right)\right]\bigg\},
\end{align}
where a new interference term is present. 
The results are shown in Fig.~\ref{fig:Fano}, with $F$  as a function of $E$ and $\theta$.
When no states are available  at the given energy, the transport is noiseless (blocked) and $F=1$, whereas when resonance with the FF is reached, $F$ exhibits a dip, signaling transport through the wire. In correspondence of the 
degeneracy point $\{E\simeq0,\theta\simeq\pi\}$ where $T_\uparrow\simeq1$, $F$ approaches zero, signaling again noiseless (perfect) transport.
%
%
%
In terms of eigenmodes $\{n\}$, indeed, $F$ reduces to $F=\sum_n T_n (1-T_n)/(\sum_n T_n)$, and $F=0$ is revealing that one eigenmode is perfectly transmitted and the other mode is completely blocked.

{\it{Experimental feasibility.~}}The transport measurements  we propose are within reach of present experimental techniques. 
Given that the most challenging ingredient is the creation of a strong enough spatially modulated field, and the  need to use large-$g$-factor semiconductors 
in order to achieve sizable Zeeman couplings, we are proposing to use large-$g$-factor InAs ($|g|\simeq15$), InGaAs ($|g|\simeq12$), and InSb wires ($|g|\simeq51$), 
or less standard materials, like InSb$_{1-x}$N$_x$ and GaAs$_{1-x}$N$_x$, with $g$-factors of several hundreds~\cite{nitrideWires}.
Moreover, our setup can exhibit effects due to strain, confinement, and interactions, which are known to significantly modify the 
$g$-factor~\cite{g_Kiselev,Strain_g_Nakaoka,InGaAs_Lin,g_Flatte,InAs_Schoenenberger,InSb_Xu}.
Taking typical magnetic fields generated by  nanomagnets, $B_n\simeq50~$mT~\cite{Karmakar_Nanomagnets}, we obtain for InSb $\Delta_n\simeq 40~\mu$eV, corresponding to $\simeq0.5~$K. 
It is convenient to choose similar values also for $\bm B$, so that the two gaps are comparable and the bound state can be observed for a large range of $\theta$ (see Fig.~\ref{fig:bulk_spectrum}), whose exact value in a measurement is not easy  to control. As explained above, the tuning 
of 
$\mu$ in the wire is achieved via an underlying gate, while the tunnel barriers are created by  gate fingers or by the wire-lead interfaces themselves. 
The \so energy in InSb wires is $\simeq50~\mu$eV, giving $2k_{\rm so}\simeq 10~\mu$m$^{-1}$, so that the above numbers would indeed place the experiment in the strong-\so regime.
The optimal transport conditions that led to the 
results of Figs.~\ref{fig:LongWire}c) and \ref{fig:LongWire}d) correspond to having $k_{l}\simeq 2$ nm$^{-1}$, probably too small to be realized in a metal lead, 
but realistic in a gated normal section of the wire serving as lead.
Corresponding estimates for $\xi_{z}(0)$ give us  $\xi_{z}(E)\simeq0.5~\mu$m, and thus a wire length of $L\simeq 1.5 ~\mu$m. 
Finally, the requirement that $B_n$ couples the $\pm2k_{\rm so}$ branches translates into having the nanomagnets separated by $2\pi/4k_{\rm so}\simeq300$ nm. 

{\it{Conclusions.~}}We have studied  the transport properties of a Rashba NW coupled to normal leads, where the interplay between 
spatially varying magnetic field, uniform field, and \so  leads to  a gap with FF states, bound to the wire ends. We have shown that there are regimes where conductance and shot noise measurements 
reveal the bound states, which can be further confirmed  by 
the $R_{\uparrow\downarrow}$ signal, if spin-resolved measurements are available.





We acknowledge discussions with R. Tiwari.
This work is supported by the Swiss NSF and the  NCCR QSIT.



\bibliographystyle{apsrev4-1} 

%
\bibliography{Transport_FF_ref} 
\end{document}